\documentstyle[epsf]{l-aa}
\def\um{\hbox{$\mu$m}}
\def\hii{H\,{\sc ii}}
\def\halpha{\hbox{H$\alpha$}}
\def\funits{\hbox{\,mJy arcsec$^{-2}$}}
\def\punits{\hbox{\,\arcsec px$^{-1}$}}
\newcount\fignumber\fignumber=1
\def\nfig{\global\advance\fignumber by 1}
\def\fignam#1{\xdef#1{\the\fignumber}}
\def\infig#1#2#3{\picplace{#1cm}}

\def\LWD{1a}		
\def\LWT{1b}		
\def\LWDSLWT{1c}	
\def\ULWD{1d}		
\def\HALWD{1e}		
\def\COLWD{1f}		

\begin{document}

\topmargin 1 in

\thesaurus{04(11.09.1; 11.09.4; 11.19.3; 13.09.1)}

\title{ISOCAM images at 6.75\,\um\ and 15\,\um: the circumnuclear
region of NGC\,4321 (M\,100) 
\thanks{Based on observations with instruments funded by ESA Member
States (especially the PI countries: France, Germany, the Netherlands
and the United Kingdom) on board ISO, an ESA project with the
participation of ISAS and NASA. } }

\author{
H.\ts Wozniak\inst{1} \and
D.\ts Friedli \inst{2,3} \and
L.\ts Martinet\inst{3} \and
D.\ts Pfenniger\inst{3}
} 
\institute{
IGRAP/Observatoire de Marseille, F-13248 Marseille Cedex 4, France
\and
D\'epartement de Physique and Observatoire du Mont~M\'egantic, 
Universit\'e Laval, Qu\'ebec, QC, G1K~7P4, Canada
\and
Observatoire de Gen\`eve, CH-1290 Sauverny, Switzerland
}

\offprints{wozniak@observatoire.cnrs-mrs.fr}

\date{Received 5 September 1997, accepted 18 September 1997}

\maketitle

\markboth{H. Wozniak et al.: ISOCAM images of NGC\,4321}{}

\begin{abstract}
We present ISOCAM images (1.5\punits) in the LW2 (6.75\,\um) and LW3
(15\,\um) broad band filters of the circumnuclear starburst of
NGC\,4321 (M\,100).  A detailed comparison with the emission at other
wavelengths ($U$, $B$, \halpha, and CO) is also presented.  The nature
and intensity of the mid-infrared emission is shown to be remarkably
different from one star forming region to the next.  For instance,
star forming regions at the ends of the nuclear bar may contain
enshrouded Wolf-Rayet stars.

\keywords{ Galaxies: individual: NGC\,4321, M\,100 -- Galaxies: ISM --
Galaxies: starburst -- Infrared: galaxies}
\end{abstract}

\section{Introduction}
This {\it Letter\/} is a first account of a global project intended to
clarify the interplay between starburst activity and the
non-axisymmetric dynamics of barred galaxies (Wozniak et al.~1997).

NGC\,4321 (M\,100), a late-type giant, moderately barred spiral galaxy
in Virgo, represents one of the best laboratories for this purpose.
Its inner region has been extensively studied at several wavelengths.
Optical (Pierce 1986), \halpha, near-infrared (Knapen et al.\
1995a,b), and CO images (Rand 1995; Sakamoto et al.\ 1995) have
revealed many intricate features associated with its mild star
formation activity in the central kpc. Several \hii\ regions lie on a
1\,kpc circumnuclear ring crossing a 8\,kpc long stellar bar.  A
nuclear bar occupies the region inside the ring (Knapen et al.\
1995a).  Its almost perfect alignment with the large-scale bar may be
coincidental since, as we know now, concentric bars can rotate at
different speeds (Friedli \& Martinet 1993).  The assumed distance to
M\,100 is 17.1\,Mpc (Freedman et al.\ 1994) which on the sky
corresponds to 83\,pc\,arcsec$^{-1}$.

The Infrared Space Observatory (ISO, Kessler et al.\ 1996) offers a
unique opportunity to explore the characteristics of the mid-infrared
(MIR) emission of this galaxy, in particular the emission coming from
``unidentified infrared emission bands'' (UIBs) and hot dust. The
actual nature of the chemical species responsible for UIBs is still
under debate, although PAH molecules are promising candidates.  The
PAHs and hot dust reemit, in the near- and mid-infrared, the absorbed
ultraviolet and visible photons (Allamandola et al.\ 1989), justifying
a multiwavelength approach to the study of star formation and dust.  A
spatial coincidence between dark dust lanes, visible on optical
images, and hot dust emission in the MIR is expected, while the
sources of exciting photons could be identified with UV images.  Also,
stars form in molecular clouds, well traced by CO emission.  However,
although a spatial association is likely to be found between CO and
hot dust emission, the precise relation remains largely unclear.

Here, we focus on the characteristics of the MIR emission as well as
on the spatial correlations and offsets with emission at other
wavelengths.

\begin{figure*}[p]
\vspace{-7truemm}
\infig{24}{figure.ps}{19.6}
\vspace{-2truecm}
\caption[1]{{\bf a} False color ISOCAM LW2 (6.75\,\um) image of
NGC\,4321.  The brightest regions are red ($\ga$\,1\funits); regions
with no data are black.  Contours run from 0.08 to 1.38\funits\ and
are spaced by 0.1\funits.  {\bf b} LW3 (15\,\um) image. The LW3
contours are scaled as for LW2.  {\bf c} LW2/LW3 colour map. Red
regions have LW2/LW3$>$1.  The huge southern black spots are an
artefact.  {\bf d} $U$ image with LW2 contours as in (a). The 0
\funits\ level is plotted as a red contour.  {\bf e} \halpha\ image
with LW2 contours as in (d).  {\bf f} CO density map with LW2 contours
as in (d). All contours are red to improve readability }
\end{figure*}

\section{ISOCAM observations and data processing}
NGC\,4321 was observed on 1996 July 8 with the ISOCAM camera (Cesarsky
et al.\ 1996) on board ISO.  We used the smallest available pixel
field of view (PFOV), i.e., 1.5\punits, which gives a
45\arcsec$\times$45\arcsec\ field of view. Two broad band filters were
used: LW2 ($\lambda_c=6.75\,\um$, $R=\lambda_c/\Delta\lambda\approx
2.25$) and LW3 ($\lambda_c=15\,\um$, $R\approx 3$).  The LW2 filter
includes UIBs emission at 6.2\,\um, 7.7\,\um\ and 8.6\,\um\ as well as
the underlying continuum. The LW3 filter collects continuum emission
of small grains as well as [NeII] (12.8\,\um) and [NeIII] (15.5\,\um)
nebular emissions.

To remove the MIR background from our images, the beamswitching mode
(AOT3) was chosen. Two empty fields 10\arcmin\ away from the galaxy
disc were selected from IRAS maps.  Each pixel on the source was
exposed twice using the exposure sequence
[sky$_1$\,$\to${}\,source\,$\to$\,sky$_2$\,$\to$\linebreak source].
Each image is composed of several hundreds of individual exposures of
2\,s each. Total exposure times of the resulting images are 1200\,s in
LW2 and 720\,s in LW3.

The raw data were reduced with CIA, the CAM Interactive Analysis
software\footnote{CIA is a joint development by the ESA Astrophysics
Division and the ISOCAM Consortium led by the ISOCAM PI, C. Cesarsky,
DSM, CEA, France.}, in a standard way (Cesarsky et al.\ 1996). We used
the IAS model (Abergel et al.\ 1997) to correct the fluxes for
transient effects. A crude estimate of the current photometric
accuracy leads to a relative error of 30\% (M.~Sauvage, private
communication).  The flux lost for a point source due to the large PSF
amounts to $8-16\%$.  The resulting S/N ratios are $\sim 40$ at
6.75\,\um\ and $\sim 10$ at 15\,\um.

The PSF beam FWHM is 3\arcsec\ at 6.75\,\um, and 6.3\arcsec\ at
15\,\um.  The LW2 resolution has been degraded down to that of the LW3
for computing the LW2/LW3 colour map. Our descriptions will mainly
refer to the LW2 image, less affected by the PSF.  We are therefore
confident of detecting features larger than 3\arcsec. Smaller details
will require the availability of improved deconvolution algorithms.

\section{Multiwavelength comparisons}
\noindent
{\bf Definitions.}  Following Knapen et al.\ (1995a), we label K1 and
K2 the two \hii\ regions close to nuclear bar ends (in coincidence
with hot spots visible in the $K$-band), whereas the two strong and
more extended \halpha\ regions, located near the extremities of the
nuclear bar minor axis, are labeled \halpha 3 and \halpha 4 (no
counterparts in $K$).

\smallskip
\noindent
{\it MIR.}  The LW2 and LW3 images are displayed in Figs.~\LWD\ and
\LWT.  Both LW2 and LW3 emissions are inhomogeneous, resulting in an
incomplete ring-like distribution.  The two bright spots near the
nuclear bar ends dominate the emission in both bands.  The K1 peak is
the brightest with $F_{6.75}$\,=\,1.39\funits\ but the emission
displays a northern extension with $F_{6.75}$\,$\approx$\,0.95\funits\
on average. For the K2 spot, $F_{6.75}$\,=\,1.25\funits. The faintest
isophote which isolates the K1 and K2 regions from the rest of the
ring is 0.85\funits. The integrated flux above this level is 71.5\,mJy
for K1 and 27.6\,mJy for K2, while their mean surface brightnesses are
similar ($\approx$\,1\funits). This is due to the larger area of the
K1 region.

Another important source is the nucleus, as bright as the K1 northern
extension ($F_{6.75}$\,=\,0.94\funits). Its integrated flux is 8\,mJy.
The mean brightness inside the circumnuclear ring, including the
nucleus, is $F_{6.75}$\,$\approx$\,0.80\funits.  No morphological
differences between the LW2 and LW3 images are perceptible without
deconvolution.  Furthermore, the bar-like feature which seems to
connect the two peaks and the nucleus remain to be confirmed with
better techniques.

Naively, one might expect that the 6.75 and 15\,\um\ emitters should
be concentrated around star forming regions, or that the MIR
brightness distribution should at least show similar steep gradients
as for \halpha.  The observed relative homogeneity may be due to the
low map resolution, as a 1.5\arcsec\ pixel covers a 125\,pc wide
region.  Alternatively, this may mean that UIBs and the underlying MIR
continuum ($5-17$\,\um) are observed everywhere.  Indeed, Mattila et
al.~(1996) detected the UIBs in the galactic disc where the radiation
field is $10^2\!-\!10^3$ lower than in \hii\ regions or planetary
nebulae. Observations of spiral galaxies (e.g.~M\,51, Sauvage et
al.~1996; NGC\,6946, Malhotra et al.\ 1996) also show continuum
emission from the whole discs.

The LW2/LW3 colour map (Fig.~\LWDSLWT) shows specific patterns: the
nuclear bar ends are dominated by the LW2 emission
(LW2/LW3$\,\approx\,$1.1--1.2) while the other two \hii\ regions are
marginally dominated by LW3 emission
(LW2/LW3\,$\approx$\,0.95). Inside the circumnuclear ring, the LW2/LW3
ratio is lower ($\approx$\,0.8). Thus, in the same object and over a
roughly similar area (circumnuclear region), we have found star
forming regions with very different MIR properties.  Two
interpretations of this ratio can be put forward.  1)~The 6.75\,\um\
excess at the K1/2 peaks could be due to the circumstellar amorphous
carbon dust present around Wolf-Rayet stars, especially WCs. ISO SWS
spectra show a large bump peaking at 7.7\,\um\ (van der Hucht et al.\
1996). A few tens of such massive stars could account for the LW2
excess over the LW3 emission.  Moreover, their bright C\,{\sc iv}
2.08\,\um\ and C\,{\sc iii} 2.11\,\um\ emission blends could be
responsible for the $K$-spots (Figer et al.\ 1997).  Thus, K1/2
starburst regions could be very young ($\la$\,$3-5$\,Myr).  2)~In
strong radiation fields, UIBs are less intense because of the
destruction of small molecules (Zavagno et al.\ 1992). This may be the
case in the \halpha 4 region which is the brightest source in the
$U$-band.  The 15\,\um\ emission could moreover be lower at the
nuclear bar ends (K1/2 regions) because the species responsible for
the continuum emission are less excited.  Together, these effects
could explain the colour map variations.

\smallskip
\noindent
{\it UV.}  A roughly 1.1\arcsec\ resolution $U$-band image, taken with
the 2.3\,m telescope at Kitt Peak Observatory, was kindly provided by
F.~Bresolin (Bresolin \& Kennicutt 1997).  It shows several bright
knots along the circumnuclear ring, but also a few peaks inside the
ring.  Figure~\ULWD\ shows that the LW2 peaks are displaced from those
in the $U$-band. In particular, the K2 peak is surrounded by
$U$-bright regions.  The only noticeable exceptions are the nucleus,
bright in $U$, LW2, LW3, and a northern overlap close to \halpha 3.

\smallskip
\noindent
{\it Visible.}  HST images, taken with broad band filters F439W ($B$),
F555W ($V$), and F702W ($R$), have been extracted from the Space
Telescope Institute archives.  The 40\arcsec\ central region of M\,100
has been observed with the PC camera at 0.046\punits.  A comparison
between HST and ISOCAM images shows that the LW2 peaks are located in,
or at the inner edge of, the dust lanes. There is also less MIR
emission in the northern and southern parts of the ring as there is
less dust.

\smallskip
\noindent
{\it \halpha.}  A roughly 3\arcsec\ resolution \halpha\ image, taken
at Mt.~M\'egantic Observatory, was kindly supplied by J.-R.~Roy and
P.~Martin.  Four main regions of star formation trace a circumnuclear
boxy ring with $a$\,$\approx$\,$500-700$\,pc, $b/a$\,$\approx$\,0.75,
and oriented at a position-angle\,$\approx$\,131\degr.  The comparison
between \halpha\ and LW2 (Fig.~\HALWD) shows that the K1 and K2 star
forming regions are spatially associated, although a little displaced,
with the LW2 peaks, while no strong counterparts to the \halpha 3 and
\halpha 4 regions are visible.  With respect to the nuclear bar
(counter-clockwise rotation), MIR maxima are trailing, whereas
\halpha\ ones leading.  This indicates that some physical properties
differ between the K1/2 and \halpha 3/4 regions.  Moreover, a
comparison between the LW3 and \halpha\ images does not confirm the
systematic connection between \hii\ regions and LW3 peaks found by
Sauvage et al.~(1996).  Indeed, the \halpha 3/4 regions have no
counterparts at 15\,\um.

\smallskip
\noindent
{\it CO.}  The 3.6\arcsec$\times$3.1\arcsec\ resolution CO density map
of Rand (1995) is used to localize the CO spiral arms and peaks. Two
molecular arms cross the bar at the location of the K1/2 regions,
while the northern and southern parts are offset toward the outer edge
of the \halpha 3/4 regions.  CO interferometric maps indicate
different physical properties in the K1/2 and \halpha 3/4 regions
(Fig.~\COLWD). On one hand, they show that the CO arms do not coincide
with the \halpha 3/4 regions and thus, they do not coincide with the
strong LW2 emission. On the other hand, the nuclear bar ends are sites
of strong CO emission, roughly correlated (i.e., inside $200-300$\,pc)
with the peaks in LW2, LW3, \halpha, $U$ and $K$-bands.

\section{Main conclusions}
1) The K1/2 regions, at the ends of the nuclear bar, show bright peaks
in LW2 and LW3, \halpha, CO, $U$ and $K$-bands located near the dust
lanes.  In these regions LW2/LW3\,$>$1, which may mean that UIB
carriers dominate the MIR luminosity. These carriers may be amorphous
carbon grains created in Wolf-Rayet stellar winds.  Knapen et
al.~(1995a,b) suggested that these regions might be young, and powered
by OB stars; while confirming this assertion, we suggest in addition
the possible presence of WCs.

\noindent
2) The H$\alpha$3/4 regions, near the ends of the nuclear bar minor
axis, are not associated with strong emission in LW2 and LW3 bands.
In these regions, the hot dust probably dominates the MIR emission,
since LW2/LW3$<$1.  The emission in \halpha, $U$ and the optical bands
is strong.  These regions are offset from the CO arms and associated
dust lanes.  They are the sites of the strongest circumnuclear
\halpha\ emission, being perhaps less obscured by dust.

\noindent
3) The centre does not show significantly strong emission in MIR
bands.  It has a low LW2/LW3 ratio ($\approx 0.8$) compared with the
surrounding ring ($\approx 0.95$) or the K1/2 regions ($\approx 1.1$).
It also emits in \halpha, $U$ and CO.

\noindent
4) There are no ``holes'' in the LW2 and LW3 surface brightness,
i.e.~some emission from UIB carriers and hot dust is observed
everywhere in the central region.  This may be either due to the low
spatial resolution ($\approx$\,250\,pc wide smoothing), or to the
existence of a significant background of MIR continuum emission.

\begin{acknowledgements} 
We are very grateful to the ISOCAM team at Saclay for many fruitful
discussions and to J.~Caplan who improved the English. We are
indebted to F.~Bresolin, P.~Martin, R.~Rand, and J.-R.~Roy for having
provided various data.  H.W.~thanks the ISO GdR (INSU) for financial
supports during his visits at Saclay.  The Swiss National Science
Foundation (FNRS) is acknowledged for its essential supports to
D.F. (``Advanced Researcher'' fellowship), and L.M. and D.P. (grant).
\end{acknowledgements}
\vspace{-5truemm}
%
{}

\end{document}